\documentclass[twocolumn,
amsmath,
amssymb,
aps,
physrev
]{revtex4-2}

\usepackage{graphicx}% Include figure files
\usepackage{dcolumn}% Align table columns on decimal point
\usepackage{bm}% bold math
\usepackage{hyperref}% add hypertext capabilities
\usepackage[mathscr]{euscript}% fancy letters
\usepackage{xcolor}

\newcommand{\ri}{{\mathrm{i}}}
\newcommand{\rd}{{\mathrm{d}}}
\newcommand{\ie}{{\it i.e.}}
\newcommand{\eg}{{\it e.g.}}
\newcommand{\tW}{{\theta_{\rm W}}}
\newcommand{\imag}{\mathfrak{Im}}
\newcommand{\del}{{\bm \nabla}}

\newcommand{\bB}{{\mathbf{B}}}
\newcommand{\bx}{{\mathbf{x}}}
\newcommand{\br}{{\mathbf{r}}}
\newcommand{\bk}{{\mathbf{k}}}
\newcommand{\bq}{{\mathbf{q}}}
\newcommand{\bb}{{\mathbf{b}}}
\newcommand{\bA}{{\mathbf{A}}}
\newcommand{\ba}{{\mathbf{a}}}
\newcommand{\bv}{{\mathbf{v}}}

\begin{document}

\title{\textbf{
Energy spectrum of magnetic fields from electroweak symmetry breaking
}}

\author{Károly Seller}
\email{Contact author: karoly.seller@uni-hamburg.de}
\affiliation{%
II. Institut für Theoretische Physik, Universität Hamburg,\\
Luruper Chaussee 149, 22761 Hamburg, Germany
}
\affiliation{Department for Theoretical Physics, ELTE Eötvös Loránd University,\\
Pázmány Péter sétány 1/A, 1117 Budapest, Hungary}
\author{Günter Sigl}
\affiliation{%
II. Institut für Theoretische Physik, Universität Hamburg,\\
Luruper Chaussee 149, 22761 Hamburg, Germany
}

\date{\today}

\begin{abstract}
    We study the magnetic fields produced in the early Universe during the electroweak symmetry breaking by considering random configurations of an inhomogeneous Higgs field.
    By exploiting the inherent randomness of the initial configurations the spectrum of the produced magnetic field is essentially analytic, which bypasses the need for costly lattice simulations.
    On the numerical side, we devise a simulation framework which results in continuous fields capable of resolving the small-scale structure of the fields that was inaccessible for the lattice-based calculation.
    Finally, by revisiting the effects of statistical isotropy and causality on the spectrum, we define general correlation functions that are then fitted to the simulation data and compared to the analytic results.
\end{abstract}

\maketitle

\section{Introduction}

The Universe today is permeated by magnetic fields, from over-dense regions of galaxies to the near-empty regions of cosmic voids.
In galaxies, the turbulent dynamics of the intergalactic plasma strengthen the magnetic field by the dynamo effect \cite{Brandenburg:2004jv,Schober_2012,Schober:2013aoa}, but non-vanishing fields exist far away from the galaxies as well, indicated by TeV blazar data \cite{Neronov:2010gir}.
It is not yet clear if the dipole-like magnetic fields of the galaxies can fully account for the field strength present in voids as the individual fields of galaxies are difficult to measure or model to sufficient accuracy (see, \eg, \cite{Unger:2023lob} for modeling the magnetic field of the Milky Way).
Due to the large-scale coherence of the magnetic fields, the author of Ref.~\cite{Vachaspati:1991nm} suggested that it could be (at least partially) a primordial relic generated by phase transitions in the early Universe.
The primary focus of investigation is naturally the electroweak phase transition (EWPT) that is predicted to have occurred in the early Universe at $T_c=(159\pm1.5)~$GeV \cite{DOnofrio:2015gop}, according to the $\Lambda$CDM model of cosmology based on the Standard Model of particle physics (SM).
The EWPT leads to the Higgs field $\Phi(\bx)$ acquiring a homogeneous non-zero vacuum expectation value (VEV) $|\Phi(\bx)|=v$, with inhomogeneities in the phases above distances of order $H^{-1}(t_0)$, at the time $t_0$ of the transition.
These inhomegeneities then lead to non-vanishing gradients of $\Phi(\bx)$, and consequently, non-vanishing magnetic fields.
For a detailed review on primordial magnetic fields, see Ref.~\cite{Vachaspati:2020blt}.

The magnetic field obtained from the inhomogeneities in the scalar field after the EWPT acts as an initial condition for the field-evolution across cosmic times described by magnetohydrodynamics \cite{Banerjee:2004df,Jedamzik:2010cy}.
This initial condition is the field that we wish to characterize in this paper, expanding on the recent results of Ref.~\cite{Vachaspati:2024vbw}.
The formation of inhomogeneities is based on topological structures and the Kibble mechanism \cite{Kibble:1976sj} (see also Ref.~\cite{Patel:2021iik} for a specific application to primordial magnetic fields).
Although the electroweak theory strictly speaking has no topological defects, it was shown in Ref.~\cite{Nambu:1977ag} that string-like configurations can exist.
These defects may form and subsequently decay during the EWPT, leaving behind the aforementioned divergence-less magnetic field.
The structure and decay of electroweak string-monopole configurations were investigated numerically in refs.~\cite{Patel:2023sfm,Patel:2023ybi}.

This paper is structured as follows.
In Sec.~\ref{sec:Magnetic_fields} we introduce the relevant concepts of the generation of magnetic fields at the EWPT.
In Sec.~\ref{sec:Magnetic_field_lattice} we briefly outline the lattice simulation, as done in Ref.~\cite{Vachaspati:2024vbw}, and show that the result is fully analytic in the limit of large lattice size.
Then in Sec.~\ref{sec:magnetic_field_continuous} we improve the simulation by employing an interpolation scheme to define a continuous magnetic field which allows for a better handling of derivatives and the small-scale structure.
The results for the correlation functions are compared to the lattice result and are fitted with empirical formulae.
Details on the derivation of analytic formulae presented in Secs.~\ref{sec:Magnetic_field_lattice} and \ref{sec:magnetic_field_continuous} are relegated to the appendices. % Apps.~\ref{app:spatial_correlation_function_on_the_lattice}--\ref{app:gen_algo}.
% Relevant formulae for the admissible correlation functions are derived in App.~\ref{app:spat_corr}, while the implications of causality are detailed in App.~\ref{app:causality}.

\section{Magnetic fields in the early Universe}
\label{sec:Magnetic_fields}

The complex scalar doublet field of the SM is conventionally written using 4 real scalar fields $\phi_i(\bx)$ as
\begin{equation}
    \label{eq:Phi_vacuum}
    \Phi(\bx)\equiv v\,\tilde\Phi(\bx)=v
    \begin{pmatrix}
        \phi_1(\bx)+\ri\phi_2(\bx) \\
        \phi_3(\bx)+\ri\phi_4(\bx)
    \end{pmatrix}\,,
\end{equation}
where the VEV of the Higgs field is defined via $|\Phi|^2=v^2$.
The vacuum manifold associated to $\Phi$ is isomorphic to $S_3$, the 3-dimensional surface of a 4-dimensional sphere.
The topologically relevant homotopy classes of $S_3$ vanish, \ie, $\pi_n(S_3)=0$ for $n=0,1,2$, signaling that the corresponding topological defects (domain walls, strings, and monopoles, respectively) are absent in the theory \cite{Kibble:1976sj} (see also Ref.~\cite{Weinberg:2012pjx} for a pedagogical review of defects).
Nevertheless, in the electroweak theory so-called ``dumbbell'' configurations are still possible: these are monopole pairs connected to each other by a $Z$-string \cite{Nambu:1977ag}.
These configurations are necessarily unstable and they decay after formation.

The electroweak monopole solution considered by Nambu may be parametrized as 
\begin{equation}
\label{eq:Phi_monopole}
    \Phi_{\rm monopole}(\bx)=f(r)
    \begin{pmatrix}
        \cos(\theta/2) \\ 
        \sin(\theta/2)\exp(\ri\varphi)
    \end{pmatrix}\,,
\end{equation}
with an overall radial profile function $f(r)$ that can be determined from minimizing the energy of the configuration (see, \eg,~\cite{Patel:2023sfm}).
The monopole lies at $r=0$ where we set $f(r=0)=0$ to avoid the singularity.
Additionally, $\Phi(\theta=\pi)\sim(0,\exp(\ri\varphi))^{\rm T}$ is a pure phase, indicating that in this direction the field is ill-defined unless we impose $\Phi(\theta=\pi)=0$.
This gives rise to a semi-infinite 1-dimensional topological defect (a string) attached to the monopole.
It is expected that monopoles pair up to form finite-size configurations, called dumbbells, with finite-length strings connecting them.

A deeper topological investigation of the electroweak theory shows that by using Hopf fibration the vacuum manifold $S_3$ can be locally deconstructed into $S_2$ with $S_1$ fibers \cite{Vachaspati:1991dz,Hindmarsh:1991jq,Gibbons:1992gt} (see also \cite{Patel:2021iik}).
Winding around $S_2$ and $S_1$ yield monopole- and string-like configurations respectively, with the overall structure of the vacuum manifold enforcing that they appear together. 
Physically, the fibration of $S_3$ can be understood via distinct gauge orbits corresponding to pure SU(2)$_{\rm L}$ and U(1)$_Y$ transformations, respectively.
In the electroweak model these orbits may originate from so-called ``semi-local'' limits, where either of the $g,g'$ gauge couplings is set to vanish \cite{Vachaspati:1991dz}, reducing the vacuum manifold to the gauge orbits of the surviving gauge group.
In reality, the gauge couplings are non-vanishing ($g,g' \neq 0$), thus the semi-local limit discussed in \cite{Vachaspati:1991dz} is not satisfied, however, mathematically the fibered structure of $S_3$ still exists.
Consequently, monopole-string configurations due to the $S_1$ and $S_2$ vacuum manifolds can be considered, with the caveat that these must necessarily be unstable. 
To connect back to the configuration \eqref{eq:Phi_monopole} discussed by Nambu, we apply the Hopf projection $\pi_{\rm Hopf}: S^3 \to S^2$ to the scalar field $\Phi \in S^3$ defined in \eqref{eq:Phi_vacuum}. 
This projection maps $\Phi$ onto the monopole configuration, $\pi_{\rm Hopf}\Phi = \Phi_{\rm monopole} \in S^2$.

Gauge invariance implies that non-vanishing gauge fields are created when the Brout-Englert-Higgs (BEH) field is spatially non-homogeneous \cite{tHooft:1974kcl,Corrigan:1975zxj}.
These non-trivial and static scalar field configurations are found by minimizing the energy functional, which amounts to setting the (spatial) covariant derivative of $\Phi$ to vanish while keeping $\Phi$ on its vacuum manifold, $|\Phi|^2=v^2$ \cite{Nambu:1977ag}.
With no independent gauge fields present, one finds a non-vanishing contribution proportional to the gradients of the BEH field,
\begin{equation}
    \label{eq:A_definition}
    \bA(\bx) = -\ri\frac{2\sin\tW}{g}\tilde\Phi^\dagger(\bx)\del\tilde\Phi(\bx)\,.
\end{equation}
Here $\tilde\Phi$ is the normalized BEH field from \eqref{eq:Phi_vacuum} and $\tW$ is the Weinberg-angle.
Consequently, a non-homogeneous scalar field configuration leads to the presence of magnetic fields \cite{Vachaspati:1991nm},
\begin{equation}
    \label{eq:B_definition}
    \bB(\bx) = -\ri \frac{2\sin\tW}{g} \del\tilde\Phi^\dagger(\bx)\times\del\tilde\Phi(\bx)\,.
\end{equation}
Obviously, if $\Phi(\bx)={\rm constant}$ then both $\bA(\bx)$ and $\bB(\bx)$ vanish, unless there are independent gauge fields present in the Universe.
In this paper, we focus on the scalar field source and assume that independent contributions are negligible.
We present results in a convention where for simplicity we set $2\sin\tW/g=1$.

\section{Magnetic field spectrum from the lattice}
\label{sec:Magnetic_field_lattice}

In the early Universe we can safely assume that due to causality, an inhomogeneous, but statistically isotropic magnetic field emerges after the electroweak phase transition.
Then, correlation functions both in direct- and in Fourier-space are used to capture the properties of this magnetic field.
The spatial correlation function for a statistically isotropic and divergence-free magnetic field $\bB(\bx)$ is (see, \eg,~\cite{Vachaspati:2020blt}):
\begin{equation}
    \langle B_i(\bx+\br)B_j(\bx)\rangle = M_{\rm N}(r)P_{ij} + M_{\rm L}(r)\hat r_i\hat r_j+\epsilon_{ijk}r_kM_{\rm H}(r),
\end{equation}
where $P_{ij}=\delta_{ij}-\hat r_i\hat r_j$, and $M_{\rm N}(r),~M_{\rm L}(r)$, and $M_{\rm H}(r)$ are the normal, longitudinal, and helical component functions, respectively.
Also, $\epsilon_{ijk}$ is the Levi-Civita symbol and the hat denotes a normalized vector, \ie, $\hat \br\equiv\br/|\br|$.
With the Fourier transform defined as
\begin{equation}
    \label{eq:FT_def}
    \bb(\bk)=\int\rd^3x~\bB(\bx)\exp(\ri\bk\cdot\bx)\,,
\end{equation}
the Fourier-space correlation function is 
\begin{equation}
\label{eq:bb_spectrum_def_general}
\begin{aligned}
    &\langle b_i(\bk)b_j(\bk')\rangle \\
    &= (2\pi)^6\delta^{(3)}(\bk-\bk')\Big[\frac{E_{\rm M}(k)}{4\pi k^2}P^{(k)}_{ij}+\ri \epsilon_{ijl}k_l\frac{H_{\rm M}(k)}{8\pi k^2}\Big]\,,
\end{aligned}
\end{equation}
where $P^{(k)}_{ij}=\delta_{ij}-\hat k_i\hat k_j$, and $E_{\rm M}(k)$ and $H_{\rm M}(k)$ is the energy- and helicity spectrum respectively.
We will be mostly interested in the non-helical part of the spectra, where
\begin{gather}
    \label{eq:spatial_corr_func_def_general}
    \langle \bB(\bx+\br)\cdot\bB(\bx)\rangle = 2M_{\rm N}(r)+M_{\rm L}(r)\,, \\
    \label{eq:fourier_corr_func_def_general}
    \langle\bb(\bk)\cdot\bb^*(\bk')\rangle = (2\pi)^6\delta^{(3)}(\bk-\bk')\frac{E_{\rm M}(k)}{2\pi k^2}\,.
\end{gather}
The non-helical correlation functions are related to each other by a spherical Hankel transform,
\begin{equation}
    \label{eq:EM_xi_relation}
    M_{\rm N}(r)+\frac{1}{2}M_{\rm L}(r)=\int_0^\infty\rd k\,E_{\rm M}(k)j_0(kr)\,,
\end{equation}
where $j_0(x)\equiv \sin(x)/x$ is the spherical Bessel function of degree 0.
In the following we shall refer to \eqref{eq:spatial_corr_func_def_general} simply as the correlation function and to \eqref{eq:fourier_corr_func_def_general} as the power spectrum.

In ref.~\cite{Durrer:2003ja} the authors showed that the energy spectrum $E_{\rm M}(k)$ should scale at least as $k^4$ for small wave-numbers in order to be compatible with causality.
As the authors pointed out, the expected energy spectrum is then of the form $E_{\rm M}(k)\propto k^4\big[1+\mathcal{O}\big((kL)^2\big)\big]$, for some characteristic distance scale $L$.
Causality in this context means that the spatial correlation function in \eqref{eq:spatial_corr_func_def_general} vanishes (or is exponentially small) for sufficiently large distances.
In the early Universe the causality argument can be applied, as fields at spatial distances larger than the Hubble radius at that time are necessarily uncorrelated.

\subsection{Setup for the lattice calculation}

Following ref.~\cite{Vachaspati:2024vbw}, we devise a lattice simulation where the lattice spacing $\delta x$ is proportional to the Hubble radius at the time of the EWPT, $\delta x\sim H^{-1}(t_0)$ \footnote{Alternatively, one can keep $\delta x$ as a free parameter and think of it as the correlation length of the BEH field after the EWPT. Depending on the specific particle physics model at hand, $\delta x$ may take any value up to a maximum at the Hubble radius, which sets a causality bound.}.
Consequently, the values of the scalar field on different lattice sites are taken independently at random.
Using the Hopf-parametrization of the Higgs field on its vacuum manifold,
\begin{equation}
\label{eq:phi_parameters}
    \tilde\Phi(\bx_{ijk})=
    \begin{pmatrix}
        \cos\alpha_{ijk} \exp(\ri\beta_{ijk}) \\
        \sin\alpha_{ijk} \exp(\ri\gamma_{ijk})
    \end{pmatrix}\,,
\end{equation}
the random configuration of the scalar field may be realized by $\beta,\gamma\in[0,2\pi)$ and $2u\equiv\cos(2\alpha)\in[-1,1]$ being chosen uniformly at random from their respective ranges, independently at each lattice site.

The gauge fields, as defined in eq.~\eqref{eq:A_definition}, may be thought of as ``links'' connecting neighboring lattice sites, a usual description in lattice field theory.
It is then useful to define the lattice link variable
\begin{equation}
\label{eq:link_variable_def}
\begin{aligned}
    T_{a}(\bx)&=\imag \big[\tilde\Phi^\dagger(\bx)\tilde\Phi(\bx+\delta x\,\hat {\bf a})\big]
    \\&\equiv
    \raisebox{-0.2cm}{\includegraphics[width=0.3\linewidth]{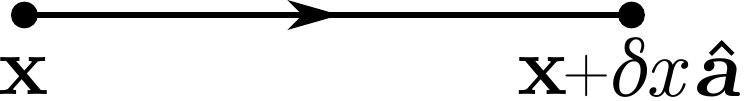}}
\end{aligned}
\end{equation}
where $a=\pm x,\pm y,\pm z$ indicates the direction of the link.
Obviously, $T_a(\bx)=-T_{-a}(\bx+\delta x\,\hat {\bf a})$, \ie, anti-parallel links between the same sites differ in sign.
By definition, $T_a(\bx)/\delta x$ gives the component of the gauge field parallel to $\hat \ba$, at the middle-point of the link.

With the gauge fields defined on the links between lattice sites, the magnetic field is assigned to a plaquette center via the loop integral
\begin{equation}
    \label{eq:Maxwell_B_def}
    \bB(\bx^{p})\delta x^2=\hat {\bf p}\oint_{\partial p}\mathbf{d\ell}\cdot\bA(\bx)\,.
\end{equation}
Here $\hat {\bf p}$ indicates the normal vector of the plaquette $p$ on whose boundary $\partial p$ we perform the line integral, and $\bx^{p}$ is the center of the plaquette $p$.
On the lattice, the integral is discretized to a sum of 4 link variables along the plaquette boundary, \eg,
\begin{equation}
\label{eq:Bx_def}
\begin{aligned}
    B_x(\bx^{p}_{i,j,k})&\delta x^2 =
    \raisebox{-1cm}{\includegraphics[width=0.3\linewidth]{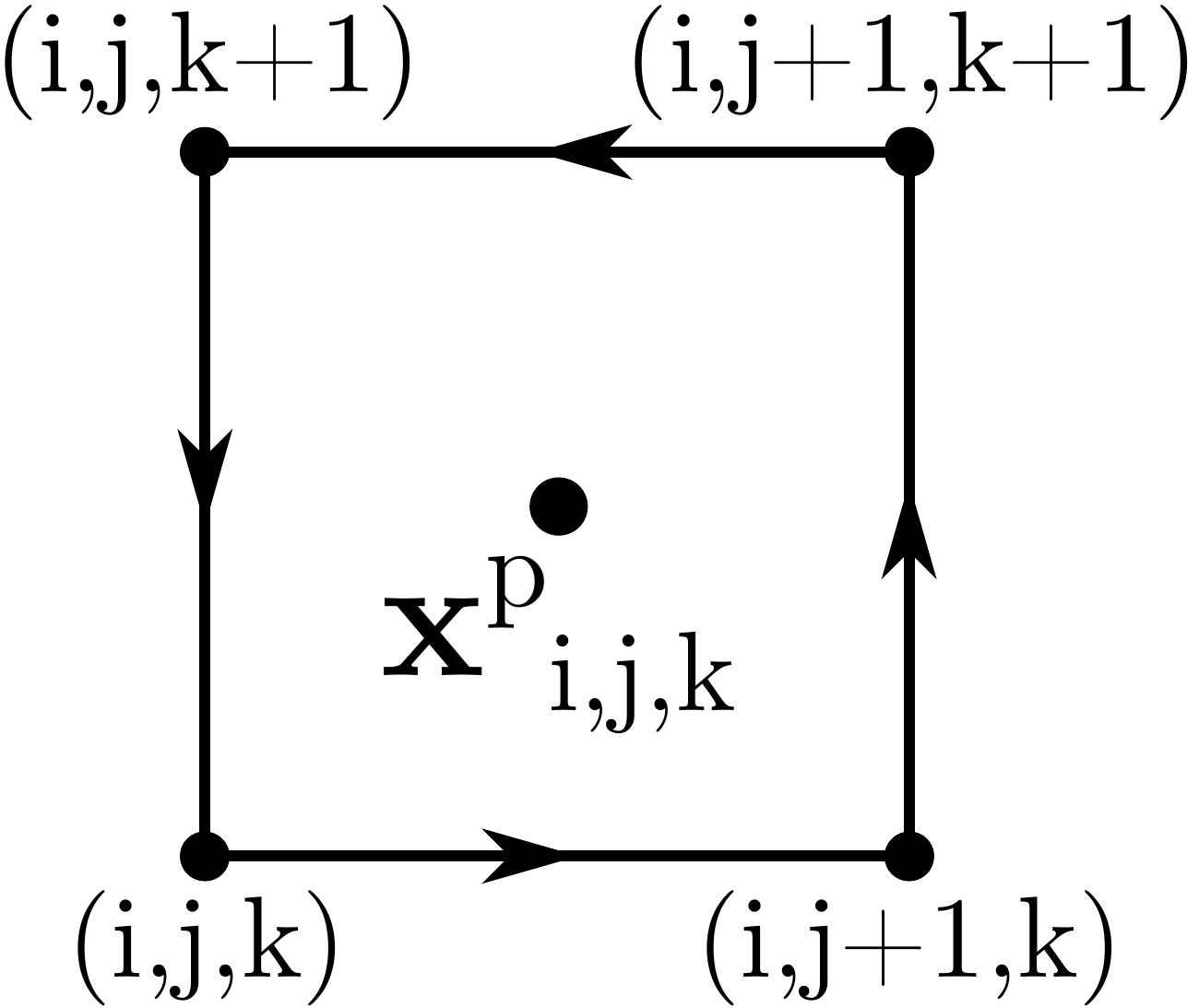}}\\
    &= T_y(\bx_{i,j,k}) + T_z(\bx_{i,j+1,k}) \\
    &\quad+T_{-y}(\bx_{i,j+1,k+1}) + T_{-z}(\bx_{i,j,k+1})\,,
\end{aligned}
\end{equation}
and similarly for the $B_y$ and $B_z$ components.
Note that each component of $\bB(\bx)$ depends only on the links within the respective orthogonal planes.

To get the power spectrum \eqref{eq:fourier_corr_func_def_general}, we perform a discrete Fourier transformation (DFT) with conventions as defined in \eqref{eq:FT_def}, and use periodic boundary conditions (PBC).
In principle other boundary conditions could also be used as we do not expect any finite correlations on distances of order the lattice size, however, PBC is natural to use with DFT.
The DFT is defined for discrete wave-numbers $\bk_{i,j,k}=k_1\,(i,j,k)^{\rm T}$ where the smallest non-zero wave-number is $k_1=2\pi/L$, with $L=N\,\delta x$ being the (linear) size of the lattice.
The largest wave-number is $k_{\rm max} \equiv \sqrt{3}k_{\rm Ny} = \sqrt{3}\pi/\delta x$, where the Nyquist frequency $k_{\rm Ny}$ is conventionally defined in 1 dimension.

The energy spectrum on the lattice for statistically isotropic fields is obtained from \eqref{eq:fourier_corr_func_def_general} with $V=(2\pi)^3\delta^{(3)}({\bf 0})$ being the lattice volume,
\begin{equation}
    \label{eq:energy_spectrum_def}
    E_{\rm M}(k)=\frac{k^2\langle|\bb(\bk)|^2\rangle}{(2\pi)^2 V}\,.
\end{equation}
We note that the common lattice continuum limit of sending $\delta x\to 0$ with keeping $L=N\delta x={\rm const.}$ is not used here as the lattice spacing $\delta x\sim H^{-1}(t_0)$ has a physical meaning and is necessarily finite.
In fact, we expect that since $\delta x$ is the {\it only} physically relevant scale in the calculation, the results for the spectra should principally depend on it.

\subsection{Analytic limit of the magnetic spectrum}

The stochastic nature of the setup described above provides an avenue to discuss the energy spectrum analytically in the limit where the lattice has sufficiently many points, in practice $N\gg 10$.
The result obtained this way yields an averaged energy spectrum over the possible initial scalar field configurations.
It is not a single, specific realization of the field, as would be the case in our Universe, nonetheless it provides the clear scaling of the spectrum and possibly allows for the discussion of further features, such as the spectral peak.

Our approach to deriving the power spectrum of the magnetic field is through the correlation function $\xi(\br)$ defined in \eqref{eq:spatial_corr_func_def_general}.
Using the ergodic principle, the average over configurations is instead written as a volume average,
\begin{equation}
    \label{eq:autocorrelation_def}
    \xi(\br) = \frac{1}{V}\int\rd^3x\,\bB(\bx)\cdot\bB(\bx+\br)\,.
\end{equation}
On the lattice, \eqref{eq:autocorrelation_def} reduces to an average of products of any two links at a given distance from each other.
As shown in App.~\ref{app:spatial_correlation_function_on_the_lattice}, this average vanishes for all products except for the case when a link is multiplied with itself.
For example,
\begin{subequations}
\label{eq:Bx2_example_pictorial}
\begin{align}
    \langle& B_x(\bx^p_{i,j-1,k})B_x(\bx^p_{i,j,k})\rangle=\\
    &=\raisebox{-1.2cm}{\includegraphics[width=0.65\linewidth]{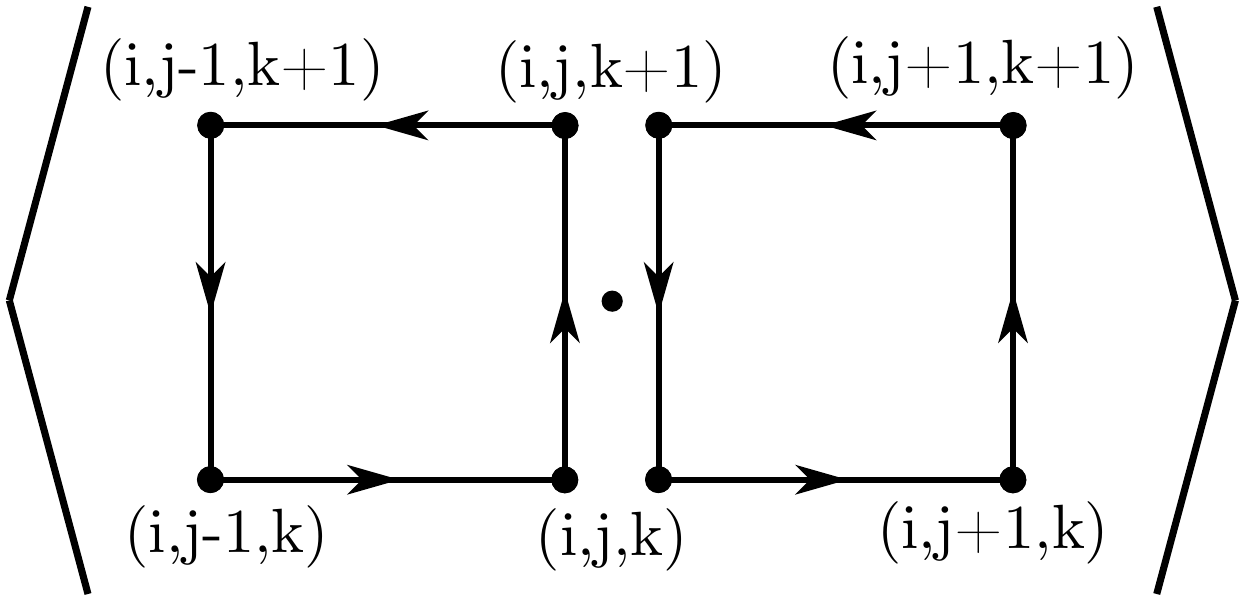}} \\
    &=\raisebox{-0.2cm}{\includegraphics[width=0.5\linewidth]{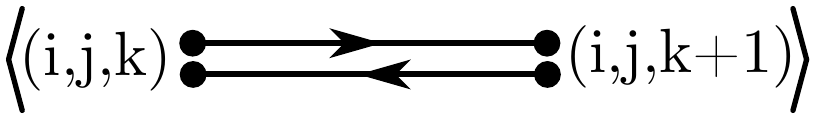}} \\
    &=\langle T_z(\bx_{i,j,k})T_{-z}(\bx_{i,j,k+1}) \rangle \equiv -\langle [T_z(\bx_{i,j,k})]^2\rangle\,.
\end{align}
\end{subequations}
In the last line we used the ``inversion'' property of the link variable introduced below \eqref{eq:link_variable_def}.
Specifically, for any squared link variable one has:
\begin{equation}
    \label{eq:Ta_exp_val}
    \langle [T_a(\bx)]^2\rangle = \frac{1}{4}\,.
\end{equation}
Recalling the definition of the magnetic field through link variables in \eqref{eq:Bx_def}, we see that such products of identical links may only appear in \eqref{eq:autocorrelation_def} if $|\br|\leq\delta x$.
Consequently, the correlation function vanishes for large distances relative to the lattice spacing $\xi(r>\delta x)=0$, thus the created field is ``causal'' and the arguments of ref.~\cite{Durrer:2003ja} apply.

On the lattice isotropy is explicitly broken.
In particular, from the autocorrelation of the magnetic field in \eqref{eq:Bx2_example_pictorial} we see that correlations vanish in non-orthogonal directions, \ie, $\langle B_i(\bx)B_i(\bx+n\delta x\,\hat {\bf e}_i)\rangle=0$ for all components (no summation for $i$ implied).
Thus, we first consider the component-wise correlations when evaluating \eqref{eq:autocorrelation_def}.
Let
\begin{equation}
   \xi(\br)=\sum_{i=1}^3\xi_i(\br)\equiv\frac{1}{N^3}\sum_{i=1}^3\sum_{\bx}B_i(\bx)B_i(\bx+\br)\,,
\end{equation}
where the second sum runs over all $N^3$ lattice points.
If the lattice is sufficiently large then we may approximately replace the sum over $\bx$ as $N^3$ times the expectation value:
\begin{subequations}
\label{eq:xi_r_fulldep}
\begin{align}
    \nonumber
    \xi(\br)&\simeq\sum_{i=1}^3 \Big[
    \delta_{\br,{\bf 0}}\big\langle [B_i(\bx)]^2\big\rangle \\ &\qquad\quad+ 
    \sum_{j=1}^3 \delta_{\br,\pm\delta x\hat{\bf e}_j} 
    \big\langle B_i(\bx)B_i(\bx\pm\delta x\hat{\bf e}_j)\big\rangle\Big] \\
    &= \frac{1}{\delta x^{4}}\sum_{i=1}^3\Big(\delta_{\br,{\bf 0}}-\frac{1}{4}\sum_{j=1}^3\delta_{\br,\pm\delta x\hat{\bf e}_j}(1-\delta_{ij})\Big)\,,
\end{align}
\end{subequations}
where an implicit summation is understood for the $\pm$ terms.
With some slight abuse of notation, we denoted the (dimensionless) Kronecker-delta for scalars, as well as for vectors as $\delta_{ab}$.

The correlation function derived in \eqref{eq:xi_r_fulldep} breaks isotropy.
Despite this, we can still define an explicitly isotropic correlation function that, contrary to $\xi(\br)$ above, is only a function of the distance $r\equiv|\br|$.
We achieve this by averaging over the angular dependence of $\xi(\br)$.
First, we introduce the set of lattice vectors $\br$ whose norm is a given value $r$, 
\begin{equation}
\label{eq:P_def}
    \mathscr{P}(r)=\big\{\br\,\big|\,|\br|=r;\,\br\in {\rm lattice}\big\}\,.
\end{equation}
From \eqref{eq:xi_r_fulldep}, with using the expectation value for the magnetic fields as in \eqref{eq:Bx2_example_pictorial} and \eqref{eq:Ta_exp_val}, the isotropic correlation function becomes
\begin{equation}
    \label{eq:corr_func_simple_lat}
    \xi_{\rm iso}(r)=\frac{3}{\delta x^4|\mathscr{P}(r)|}\big(\delta_{r,0}-\delta_{r,\delta x}\big)=\frac{3}{\delta x^4}\Big(\delta_{r,0}-\frac{1}{6}\delta_{r,\delta x}\Big)\,,
\end{equation}
where $|\mathscr{P}(r)|$ is the cardinality of the set $\mathscr{P}(r)$.
As $\xi_{\rm iso}(r)$ is explicitly defined on the lattice, it is not a continuous function of $r$, rather it is only defined for discrete distances allowed on the lattice.
However, the continuous ``limit'' of $\xi_{\rm iso}(r)$ may be obtained by recognizing that in the continuum $\mathscr{P}(r)\to 4\pi (r/\delta x)^2$ and $\delta_{r,a}\to\,\delta x\,\delta(r-a)$.
A more detailed derivation of \eqref{eq:corr_func_simple_lat} is presented in App.~\ref{app:spatial_correlation_function_on_the_lattice}.

The power spectrum on the lattice is the DFT of the spatial correlation function,
\begin{equation}
    \label{eq:bk_DFT}
    |\bb(\bk)|^2 = V\delta x^3\sum_{\br}\xi(\br)\exp(\ri\bk\cdot\br)\,,
\end{equation}
where $\bk$ are discrete wave-numbers as defined by the DFT and $\xi(\br)$ is given in \eqref{eq:xi_r_fulldep}.
As the correlation function (even in the non-averaged case) satisfies $\xi(\br)=\xi(-\br)$, only the real part of the complex exponential (cosine function) contributes.
It follows then that $|\bb(\bk)|^2$, and consequently $E_{\rm M}(k)\propto k^2\langle|\bb(\bk)|^2\rangle$ are even functions of the wave-number $k$.

The correlation function vanishes for large distances, thus the sum in \eqref{eq:bk_DFT} reduces to the sum of contributions from $\br=\mathbf{0}$ and $\br=\pm\delta x\,\mathbf{e}_i$.
Finally,
\begin{equation}
\label{eq:b2_lattice}
\begin{aligned}
    \frac{|\bb(\bk)|^2}{\delta x^2 N^3} = \sum_{i=1}^3 \Big[1-\cos\big(k_i\delta x\big)\Big]\,,
\end{aligned}
\end{equation}
where $k_i=2\pi m_i/(N\delta x)$ with $m_i\in[0,N/2]$.
The right-hand side of \eqref{eq:b2_lattice} depends separately on the components of the wave-number vector (although in an interchangeable way), showing the explicit breaking of isotropy on the lattice inherited from $\xi(\br)$.

For small wave-numbers (\ie, $m_i\ll N$) we may expand the cosine functions in \eqref{eq:b2_lattice} as $\cos(x)\simeq1-x^2/2$.
The constant term cancels and the leading non-vanishing term is found to be proportional to the square of the wave-number $m_x^2+m_y^2+m_z^2\equiv k^2/k_1^2$.
Thus at leading order the power spectrum is isotropic.
The power spectrum is:
\begin{equation}
    \label{eq:b2_scaling}
    \langle|\bb(k)|^2\rangle = \frac{N}{2}\delta x^2(kL)^2\Big[1+\mathcal{O}\big((kL)^2\big)\Big]\,.
\end{equation}
Then from \eqref{eq:energy_spectrum_def} the energy spectrum for small wave-numbers becomes 
\begin{equation}
    \label{eq:EM_small_k_scaling_simple}
    8\pi^2\frac{E_{\rm M}(k)}{k^4}=\delta x\big[1 +  \mathcal{O}\big((kL)^2\big)\big]\,.
\end{equation}
This finding is consistent with the causality argument of ref.~\cite{Durrer:2003ja}, and completes our proof of the initial magnetic energy spectrum after the EWPT scaling with $k^4$ for small wave-numbers.
The expression for $E_{\rm M}(k)$ in \eqref{eq:EM_small_k_scaling_simple} is independent of $N$ or the volume $V$, and only depends on the one physical length scale of the problem, the correlation length $\delta x\sim H^{-1}(t_0)$.

Eq.~\eqref{eq:b2_lattice} can also be used to calculate the full energy spectrum with no expansion in $k/k_1$.
However it has to be kept in mind that for larger values of $k$ the spectrum will suffer from discretization effects.
Additionally, while \eqref{eq:b2_lattice} explicitly breaks isotropy beyond leading order, one can still average over the directions of the wave-number vectors, as explained in App.~\ref{app:expansion_of_the_energy_spectrum}.
We find for $k\leq k_{\rm Ny}$ that the angular-averaged lattice energy spectrum is
\begin{equation}
    8\pi^2 \frac{E_{\rm M}(k)}{k^4}=\frac{6}{k^2\delta x}\big[1-j_0(k\delta x)\big]\,.
\end{equation}
Since $\rd E_{\rm M}(k)/\rd k\neq 0$ for any $k>0$, this spectrum does not have a spectral peak below the Nyquist wave-number.

While the approach presented in this section is useful for predicting the overall scaling of the energy spectrum for small wave-numbers, it is insufficient to help in determining additional features of the spectrum due to the crudely defined derivatives and low spatial resolution.
This issue may be alleviated by introducing more carefully defined derivatives or a more densely populated lattice.
The latter is excluded by the algorithmic approach, as we cannot generate scalar field values independently at nearby ($r\lesssim H^{-1}(t_0)$) spatial points.
In Ref.~\cite{Vachaspati:2024vbw} the authors used a setup where the magnetic field on the lattice sites was defined by averaging over those calculated at the center points of neighboring plaquettes.
This coarse-grained approach is equivalent to using \eqref{eq:Maxwell_B_def} with a 4-plaquette averaging (4PA).
In fact, the coarse-grained method can be freely generalized to $n^2$PA methods, although the impact of going beyond 4PA is negligible, as outlined in app.~\ref{app:gen_algo}.

\begin{figure}
    \centering
    \includegraphics[width=\linewidth]{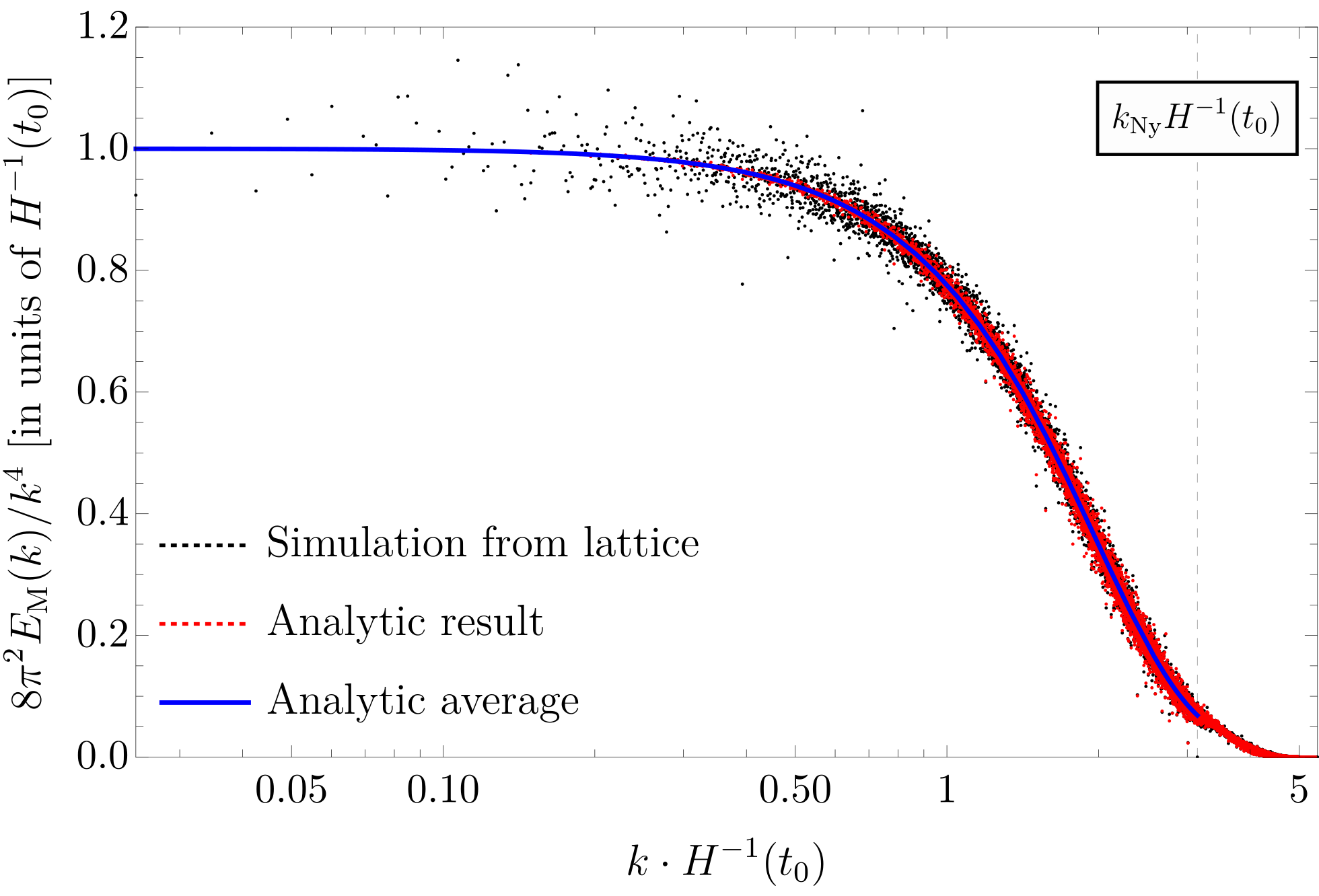}
    \caption{Energy spectrum calculated in the 4PA case on an $N=256$ lattice with averaging over 10 realizations ({\it black dots}) and from the analytic calculation ({\it red dots}).
    The analytic average written in terms of spherical Bessel functions is overlaid on the numerical results ({\it blue}).
    For small wave-numbers the spectrum is seen to be $\propto k^4$.
    }
    \label{fig:Energy_spectrum_compare_4PA}
\end{figure}

Correlations for the 4PA magnetic field extend to greater distances than in the original calculation, as the larger plaquettes will share common links with each other up to $r_{\rm max}=\sqrt{5}\,\delta x$.
The resulting isotropic spatial correlation function (see, apps.~\ref{app:spatial_correlation_function_on_the_lattice} and \ref{app:gen_algo}) is
\begin{equation}
    \xi_{\rm iso}^{\rm (4PA)}(r)
    =\frac{2\cdot 3}{(2\delta x)^4|\mathscr{P}(r)|}\Big[\delta_{r,0} + \delta_{r,\delta x} - \delta_{r,2\delta x} - \delta_{r,\sqrt{5}\delta x}\Big]\,,
\end{equation}
where $|\mathscr{P}(\delta x)|=|\mathscr{P}(2\delta x)|=6$ and $|\mathscr{P}(\sqrt{5}\delta x)|=24.$
The corresponding averaged energy spectrum is derived in app.~\ref{app:expansion_of_the_energy_spectrum} and is given by
\begin{equation}
\label{eq:b2_analytic_lattice_4PA_main}
\begin{aligned}
    &8\pi^2 \frac{E_{\rm M,4PA}(k)}{k^4} \\
    &= \frac{3}{4k^2\delta x} \big(1+j_0(k\delta x)-j_0(2k\delta x)-j_0(\sqrt{5}k\delta x)\big)\,.
\end{aligned}
\end{equation}
In the 4PA approach (i) the leading order result for the energy spectrum at small wave-numbers matches that of \eqref{eq:b2_scaling} and (ii) the spectral peak exists.
When the analytic spectrum is compared to numerical simulations of the sort presented in Ref.~\cite{Vachaspati:2024vbw}, we find perfect agreement.
In Fig.~\ref{fig:Energy_spectrum_compare_4PA} we present a reproduction of their numerical results overlaid with our analytic formulae.
For small values of $k$ the lattice spectrum is noisy because of low statistics, whereas for larger values the simulated values (black dots) closely match those of the analytic limit (red dots) obtained from \eqref{eq:b2_lattice}.

The 4PA introduced features that one expects from the energy spectrum, however the physical interpretation of this setup becomes questionable as correlations now exist on super-horizon scales, up to $r=\sqrt{5}H^{-1}(t_0)$.
In the next section we perform a refined lattice simulation where we use continuous interpolation to be able to define the scalar field at all spatial points based on an initial random configuration similar to that used in this section.
The interpolation also allows for analytic derivatives everywhere, thus the magnetic field is available locally as the curl of the vector potential.
We do not expect differences for small wave-numbers (large distances) as this setup is also inherently causal, but we do expect a more precise encapsulation of smaller scale ($r\lesssim \delta x$) behavior of the energy spectrum.

\section{Magnetic field spectrum from continuous fields}
\label{sec:magnetic_field_continuous}

In this section our goal is to use non-linear interpolation to find the magnetic field at arbitrary spatial points within the initial lattice volume.
To ensure that we control causality through the correlation length, we will first define a lattice of initial random seeds for the scalar field with physical size $L^3$. 
Then the lattice spacing of the seed-lattice sets the scale of the correlation length through $\delta x=L/N_{\rm seed}\sim H^{-1}(t_0)$.
Physically, the seeds correspond to the scalar fields within the initial domains of the broken phase after phase transition.
We then use interpolation to define the scalar field at all points, from where the gauge fields are available directly from \eqref{eq:A_definition} and the magnetic field is obtained from $\bB(\bx)=\mathbf{\nabla}\times\bA(\bx)$.

To ensure that the scalar field remains on its vacuum manifold $S_3$ everywhere, instead of interpolating the scalar field components, we directly interpolate the Hopf-parameters $u$ and $\beta,\gamma$.
With the parameters of the scalar field defined for any $\bx$, we can evaluate the magnetic field directly as
\begin{equation}
    \label{eq:parametric_B_field_from_curl}
    \bB(\bx)=\nabla\zeta(\bx)\times\nabla u(\bx)\,,
\end{equation}
where we introduced the relative phase between the components of the scalar field, $\zeta(\bx)=\gamma(\bx)-\beta(\bx)$.

Interpolation of phases in more than 1 dimension runs into phase ambiguities arising from the incapability of the interpolation to grasp the inherent periodicity of the phases.
This may be partially alleviated by so-called phase unwrapping algorithms, that shift the phases by integer multiples of $2\pi$ to get a phase map where all neighboring points are within $\pm\pi$ distance from each other.
In 1 dimension this is uniquely achievable, however, in higher dimensional spaces a complete unwrapping is in general impossible.
In these cases the unwrapping algorithms work by minimizing the number of points where the phases jump by more than $\pm\pi$ (see ref.~\cite{Abdul-Rahman:07} for 3D phase unwrapping).
The lack of a complete unwrapping is caused by topological defects such as those leading to monopoles or strings, \ie, path-dependent assignment of unwrapped phases.
Since we wish to describe the magnetic field that is divergence-free (\ie, the formulae for the spectrum apply), we assume that all topological configurations have decayed at the time of investigation.
Then the simplest approach is to ignore the periodicity of the phases and directly interpolate them.
This ensures continuously varying phases that are random at distances of order $H^{-1}(t_0)$ and thus a single-valued inhomogeneous scalar field without any topological defects.

One concern not raised so far about the lattice simulation of the previous section is in its inherent bias in setting up a regular cubic lattice of seeds.
The underlying cubic structure will be non-trivially present in all derived quantities, causing unphysical features in the correlation function.
We aim to resolve this by randomizing our grid by shifting the physical coordinate of each lattice site by a random 3-vector.
These shifts are independently performed on all lattice sites such that there is no overlap, and a minimum distance between lattice sites is enforced.
By design, the average distance between lattice sites remains the original lattice spacing, \ie, the desired correlation length.
This approach leads to standard interpolating techniques, such as cubic splines, being unusable due to the non-uniformly defined grid.
Instead, we use radial basis function (RBF) interpolation that also allows for the interpolation of non-regular grids.
Similarly to splines, the derivative of the RBF interpolated function is also analytic.
The base RBF interpolation is implemented in \eg, \verb|python|'s \verb|scipy| package \cite{Wahba:1990}.

\begin{figure}
    \centering
    \includegraphics[width=\linewidth]{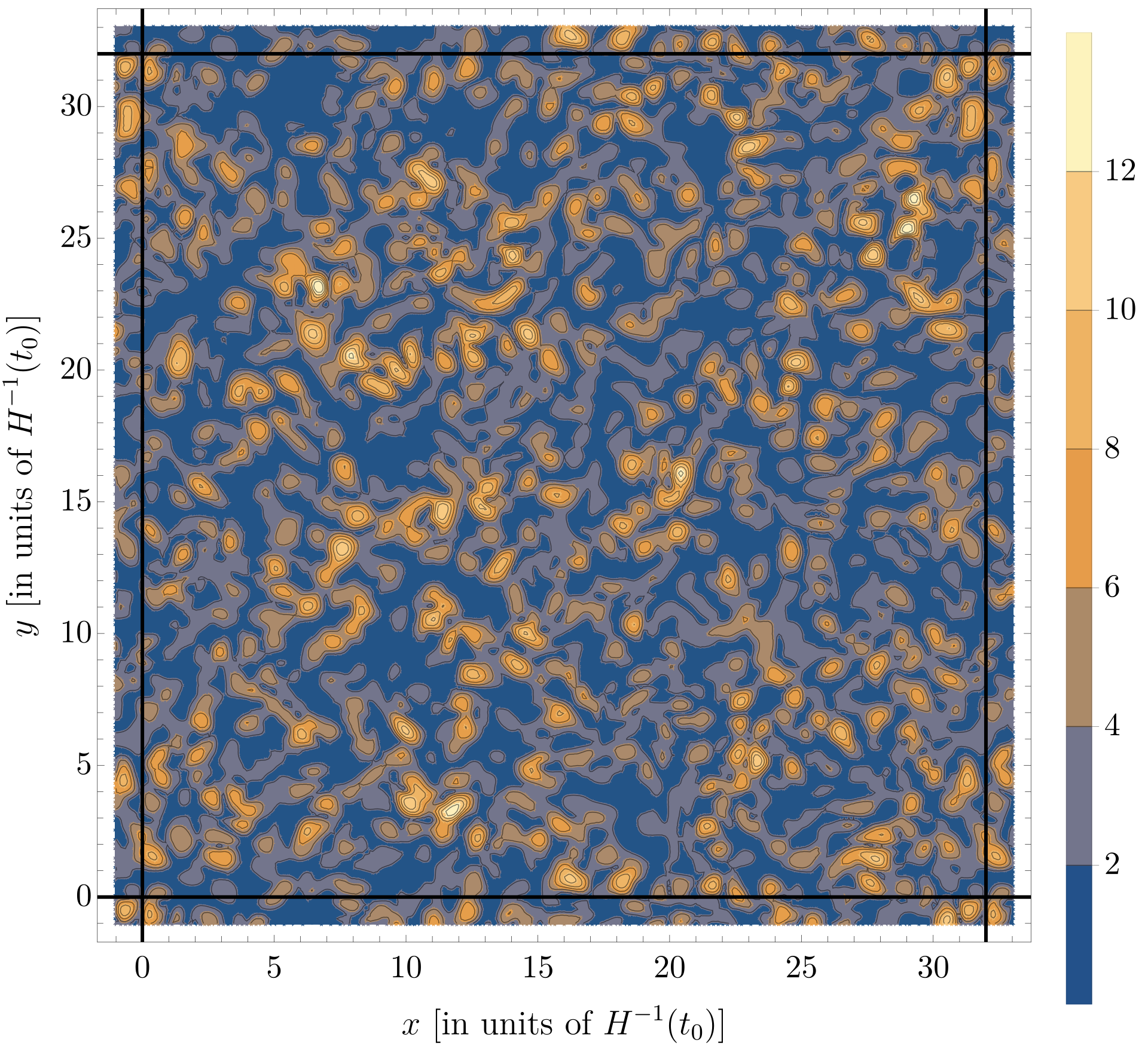}
    \caption{An example slice of the magnitude of the magnetic field for an initial lattice of size $N_{\rm seed}=32$ (periodically extended beyond the boundaries indicated by black lines).
    The magnitude scale on the right-hand side is understood in units of $H^2(t_0)$.}
    \label{fig:Bfield_example}
\end{figure}

The issue with interpolating the parameters of the scalar field is that the various ranges for the parameters are generally not respected by the interpolated function.
This can be easily fixed by using a clamp on the parameter functions.
In practice, we used a logarithmic clamp with a sharpness-controlling parameter $\alpha$ to restrict the values of a function $f(\bx)$ to the range $[a,b]$,
\begin{equation}
    f_{[a,b]}(\bx;\alpha)=a+\frac{1}{\alpha}\log\Bigg[\frac{1+\exp\big(\alpha(f(\bx)-a)\big)}{1+\exp\big(\alpha(f(\bx)-b)\big)}\Bigg]\,.
\end{equation}
The clamping was used for the parameter $u$ where respecting the proper boundaries was crucial to keep $\tilde\Phi(\bx)$ properly normalized.
For the phases, the clamp is not necessary as the sine and cosine functions will still be continuously defined for all values above or below the boundaries.
In the numerical results presented in this paper we have used $\alpha=15$.
The PBC for the lattice volume can be enforced during interpolation by periodically extending the lattice in each direction by $N_{\rm pad}$ points.
With sufficiently many points (in practice a few is enough as interpolation only depends on close-by points) the function values and derivatives can be made to match on the boundary (see Fig.~\ref{fig:Bfield_example}).

Once the interpolated magnetic field is defined, we introduce a new, regular grid with $N_{\rm lattice}>N_{\rm seed}$, where the magnetic field is re-sampled for the DFT algorithm.
Then the energy spectrum is obtained identically to the method described in Sec.~\ref{sec:Magnetic_field_lattice}.
The relevant parameter of the calculation is the ratio $N_{\rm lattice}/N_{\rm seed}$ that tells us on average how many new points lie between the initially set up correlation length.
The physical size of the system is taken as $L=N_{\rm seed} H^{-1}(t_0)$, thus the minimal and maximal wave-numbers are $k_1H_0^{-1}(t_0)=2\pi/N_{\rm seed}$ and $k_{\rm Ny}H_0^{-1}(t_0)=\pi N_{\rm lattice}/N_{\rm seed}$.
An example of a spatial slice of the magnitude of the magnetic field calculated with $N_{\rm seed}=32$ is shown in Fig.~\ref{fig:Bfield_example}.

The expected energy spectrum is an even polynomial in $k$ \cite{Durrer:2003ja}, from which we factorize an exponential, Gaussian-type cutoff for large $k$.
A general consequence of causality and isotropy is that the energy spectrum depends only on even powers of $k$ (see App.~\ref{app:causality}).
The assumed exponential suppression at large $k$ occurs only if the correlation function decays as $\exp(-ar^2)$ or faster; a decay of $\exp(-ar)$ would lead only to a power-law decay in $k$.
For even values of $n$ we use the following empirical formula,
\begin{equation}
    \label{eq:power_spectrum}
    \mathcal{E}^{(n)}(k;\{b_{a}\},\sigma)\propto k^4\Big(1+\sum_{a=1}^{n/2}b_ak^{2a}\Big)\exp\Big(-\frac{2k^2}{\sigma^2}\Big)\,.
\end{equation}
The parameters $b_a$ and $\sigma$ are obtained from fitting to the simulated spectra.
The physical significance of the latter is its relation to the position of the spectral peak $k_*$. 
In particular the exponent is normalized such that for $n=0$ one has $k_*=\sigma$.
This relation no longer holds for $n>0$.

\begin{figure}
    \centering
    \includegraphics[width=\linewidth]{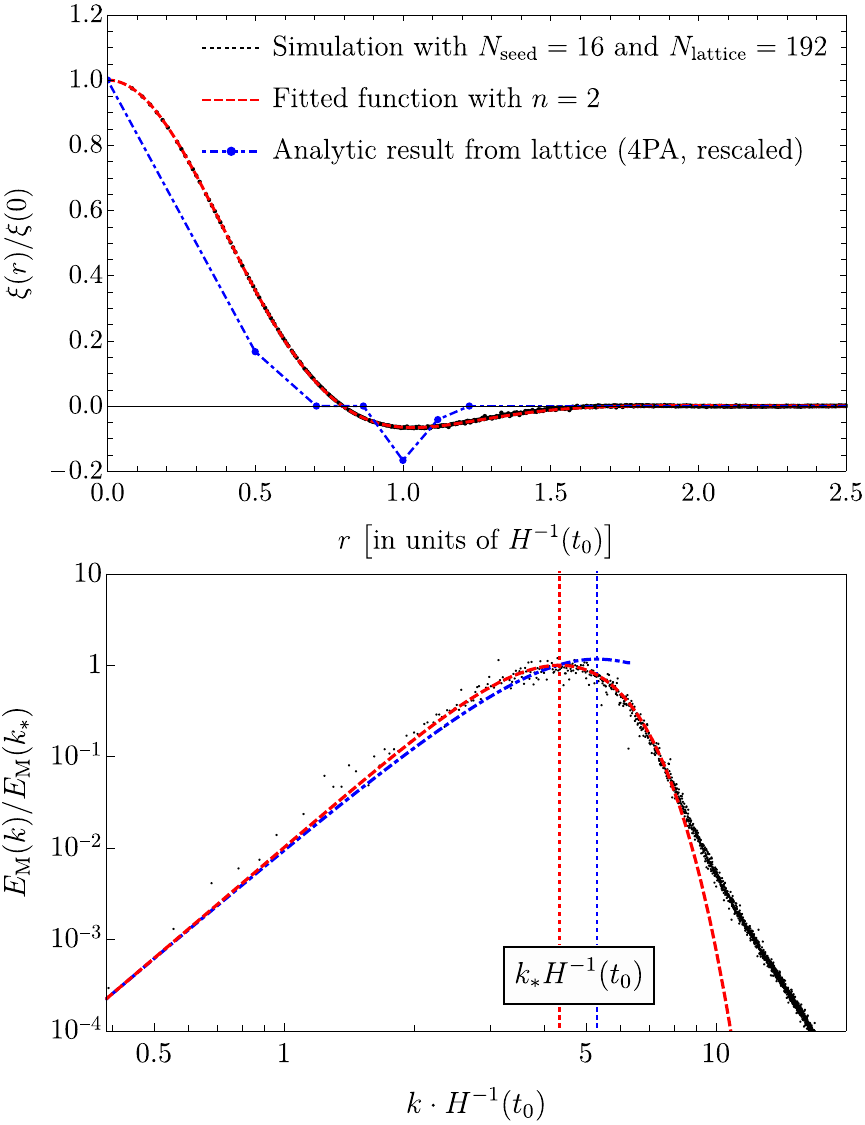}
    \caption{Spatial correlation function $\xi(r)$ ({\it top figure}) and energy spectrum $E_{\rm M}(k)$ ({\it bottom figure}) from the improved lattice simulation with $N_{\rm seed}=16$ and $N_{\rm lattice}=192$.
    The fitted functions of $\mathcal{C}^{(2)}(r)$ and $\mathcal{E}^{(2)}(k)$ ({\it red, dashed}) and the 4PA analytic functions ({\it blue, dot-dashed}) are also plotted in the respective figures as comparisons to the simulation result.
    In the bottom figure, vertical lines indicate the position of the spectral peak in the 4PA ({\it blue, dotted}) and improved simulation fit case ({\it red, dotted}).}
    \label{fig:CombinedSpectra}
\end{figure}

Using isotropy of the magnetic field, the expected autocorrelation function for the magnetic field is found by an appropriate Hankel transform of the energy spectrum,
\begin{equation}
    \label{eq:Cn_def}
    \mathcal{C}^{(n)}(r) = 2\int_0^\infty\rd k\, \mathcal{E}^{(n)}(k;\{b_{a}\},\sigma)j_0(kr)\,.
\end{equation}
This integral is analytic for energy spectra of the form given in \eqref{eq:power_spectrum}, and retains the exponential suppression for large $r$, \ie, it satisfies the causality requirement.
In particular, $\mathcal{C}^{(n)}(r)$ takes the form of an even polynomial in $r$, multiplied by an exponential suppression factor,
\begin{equation}
    \mathcal{C}^{(n)}(r)=\sum_{m=0}^{m_{\rm max}}p_mr^m\exp(-ar^2)\,.
\end{equation}
Importantly, for a general ``causal'' correlation function the polynomial is not arbitrary.
The coefficients $p_n$ have a very specific form to cancel the $k^2$ term in the corresponding energy spectrum $\mathcal{E}^{(n)}(k)$, the exponential suppression is not enough.
As shown in App.~\ref{app:spat_corr} the coefficients must satisfy 
\begin{equation}
    \sum_{m=0}^{m_{\rm max}}p_m a^{-m/2}\Gamma\Big(\frac{3+m}{2}\Big)=0\,.
\end{equation}
For the explicit formula of $\mathcal{C}^{(2)}(r)$ used in Fig.~\ref{fig:CombinedSpectra} and the formulation of ``admissible'' correlation functions, see App.~\ref{app:spat_corr}.

The simple relationship between the energy spectrum $\mathcal{E}(k)$ and the correlation function $\mathcal{C}(r)$ provides an additional avenue to discuss the scaling of the energy spectrum for small wave-numbers directly from $\mathcal{C}(r)$.
Given a causal correlation function $\mathcal{C}(r)$ that leads to an energy spectrum that scales as $\propto k^4$, the following constraints are automatically satisfied:
\begin{subequations}
\label{eq:k4_conditions}
\begin{align}
    \label{eq:k4_conditions_i}
    \text{(i)}\quad &\int_0^\infty\rd r\,r^2\mathcal{C}(r)=0\\
    \text{(ii)}\quad &\int_0^\infty\rd r\,r^4\mathcal{C}(r)\neq 0\,.
\end{align}
\end{subequations}
Condition (i) enforces that the $\propto k^2$ term cancels in $\mathcal{E}(k)$ while condition (ii) forces $\mathcal{E}(k)\propto k^4$.
Although the conditions may seem to be additional constraints for $\mathcal{C}(r)$ over that of causality (\ie, rapid decay at large separations), they follow directly from the analytic properties of the Fourier transform, as explained in App.~\ref{app:causality}.
Put more precisely, (i) is always satisfied for any causal correlation function, but (i) does not lead to a $\mathcal{O}(k^4)$ spectrum if $\mathcal{C}(r)$ is not causal.
Obviously, \eqref{eq:k4_conditions} holds for $\mathcal{C}^{(n)}(r)$ by construction, but as shown in App.~\ref{app:expansion_of_the_energy_spectrum}, they also hold for a suitable generalization of the lattice results presented in Sec.~\ref{sec:Magnetic_field_lattice}.

We show an example for the correlation function and the energy spectrum in fig.~\ref{fig:CombinedSpectra}.
In the figure, the improved simulation results (with lattice parameters $N_{\rm seed}=16$ and $N_{\rm lattice}=192$) are plotted versus the fits of the empirical functions given in \eqref{eq:power_spectrum} and \eqref{eq:Cn_def} with a fixed order of $n=2$ (higher order terms are irrelevant even if they are included in the fit).
To compare with the original lattice results, we have also plotted the analytic 4PA correlation function and energy spectrum.
In the plots, the 4PA results are rescaled in $k$, as explained below \eqref{eq:b2_analytic_lattice_4PA}.
The position of the spectral peak is somewhat below the value obtained from the original lattice simulation, here we obtained $k_*\delta x\approx 4.35$, corresponding to $\lambda_*\approx 1.44\delta x$.
As expected, a better spatial resolution shifted the location of the spectral peak closer to the wave-number corresponding to wavelengths equal to the Hubble radius at the time of EWPT.

\section{Conclusions}

In this article, we revisited the production mechanism of primordial magnetic fields at the electroweak phase transition, employing techniques from lattice field theory to estimate an average magnetic field configuration. Using analytic arguments, we demonstrated that the magnetic energy spectrum scales as $E_{\rm M}(k)\propto k^4$ at small wave numbers. In particular, we derived the lattice-averaged spectrum $E_{\rm M}(k)\propto k^2\big(1 - j_0(k\,\delta x)\big)$ for $k < k_{\rm Ny}$. We further showed that this small-$k$ scaling is universal across coarse-grained lattice simulations. These analytic results were validated against numerical simulations and found to be in full agreement, enabling one to bypass the costly numerical generation of field configurations.

In addition to these analytic results, we introduced a more realistic lattice setup designed to capture small-scale inhomogeneities of the magnetic field. This construction employs a non-linear interpolation on a non-uniform grid of initially random seed field values to generate a continuous field configuration. As a result, all quantities are defined at arbitrary spatial points, with the correlation length controlled by the average spacing of the seed values. We showed that this approach produces an energy spectrum consistent with the lattice results and proposed phenomenological expressions to model it. In particular, a low-degree polynomial with an exponential cutoff provides an accurate fit in both real and Fourier space. Finally, we presented general admissible correlation functions satisfying causality and statistical isotropy.

Our results are fully consistent with those of Ref.~\cite{Durrer:2003ja}, based on causality arguments, and with the numerical findings of Ref.~\cite{Vachaspati:2024vbw}, which served as a primary motivation for this work.

\section{Acknowledgments}

This research was supported by the Deutsche Forschungsgemeinschaft (DFG, German Research Foundation) under Germany’s Excellence Strategy– EXC 2121 “Quantum Universe”– 390833306.
This research was partially funded by Excellence Programme of the Hungarian Ministry of Culture and Innovation grant number TKP2021-NKTA-64 
and by the Hungarian Scientific Research Fund grant number PD-146527.
KS would like to thank Tanmay Vachaspati for providing the numerical code for the lattice simulation presented in ref.~\cite{Vachaspati:2024vbw} that proved useful for cross-checking the validity of the results presented here.

\appendix
\section{Spatial correlation function on the lattice}
\label{app:spatial_correlation_function_on_the_lattice}

The basic building blocks for the magnetic field are the link variables defined in \eqref{eq:link_variable_def}.
A trivial calculation shows that the expectation value of a single link vanishes due to the phases averaging to zero, $\langle T_a(\bx)\rangle=0$.
Consequently, from \eqref{eq:Bx_def} we find $\langle B_i(\bx)\rangle=0$ for the expectation value of the magnetic field at a single point.
However, the variance of the link variables, and consequently of the magnetic field, is non-zero:
\begin{equation}
    \label{eq:link_sqr_exp_val}
    \langle [T_a(\bx)]^2\rangle=\frac{1}{4} \quad \rightarrow\quad \langle[B_i(\bx)]^2\rangle=(\delta x)^{-4}\,.
\end{equation}
The second relation follows from the first by recognizing that the magnetic field squared involves 4 squared link variables. 
Note, that the expectation value of products of links connecting different sites is zero due to the independently assigned random variables at each site.

%\begin{figure}
%    \centering
%    \includegraphics[width=\linewidth]{Figs/SingleLink.pdf}
%    \caption{Non-vanishing contributions in the correlation function: the $r=0$ case ({\it left}) and one example for the $r=\delta x$ case ({\it right}).}
%    \label{fig:single_link}
%\end{figure}

The spatial correlation function, as defined in \eqref{eq:autocorrelation_def}, is discretized on the lattice, where it is a sum over the scalar product of magnetic fields at points $\bx$ and $\bx+\br$.
As the magnetic field is a sum of 4 links around a plaquette, the product of magnetic fields reduces to products of 2 link variables.
As mentioned, the expectation values of products of links vanish, except for the case in \eqref{eq:link_sqr_exp_val}.
It follows then that only $r=0$ and $r=\delta x$ may contribute to the correlation function, as the product of $\bB(\bx)$ fields calculated at larger separations do not share common links.

For $r=0$ the correlation function gives the average of the magnetic field square over the entire lattice (proportional to the magnetic energy density). 
In the large $N$ limit this is evaluated as
\begin{equation}
\begin{aligned}
    \xi_{\rm iso}(0)&=\frac{1}{N^3}\sum_{i,j,k=1}^{N} |\bB(\bx_{ijk})|^2 \simeq \frac{3}{N^2} \sum_{j,k=1}^{N} |B_x(\bx_{ijk})|^2 \,.
\end{aligned}
\end{equation}
In the second equality, due to statistical isotropy, we could make the replacement $|\bB|^2\to 3B_x^2$ inside the summation.
Moreover, as $B_x(\bx)$ only depends on links in the $y-z$ plane, the sum over the $x$ coordinates becomes trivial, $\sum_{i,j,k}\to N\sum_{j,k}$. 
What remains is a sum for squared link variables within a single sheet.
In particular, each squared link will appear twice as each link borders exactly two plaquettes in the plane of the sheet.
The total number of links within a sheet of size $N\times N$ with PBC is $N_{\rm links}=2N^2$, thus
\begin{equation}
    \label{eq:corr_func_0}
    \xi_{\rm iso}(0) \simeq \frac{3}{N^2\delta x^4}\cdot\langle|T_a(\bx)|^2\rangle\cdot2 N_{\rm links}=\frac{3}{\delta x^4}\,.
\end{equation}

For $r=\delta x$, we use that neighboring plaquettes share a single anti-parallel link.
There are in total 6 such neighbors from which only 4 share a common link (those plaquettes lying within the same plane).
As each of the latter share a single anti-parallel link as opposed to the 4 parallel links used in $\xi_{\rm iso}(0)$, they each contribute $-\xi_{\rm iso}(0)/4$.
The isotropic correlation function for $r=\delta x$ is thus
\begin{equation}
    \label{eq:corr_func_dx}
    \xi_{\rm iso}(\delta x)=-\frac{\xi_{\rm iso}(0)}{4}\cdot\frac{4}{6}\simeq-\frac{1}{2\delta x^4}\,.
\end{equation}
The magnetic field is anti-correlated at distances comparable to the correlation length.

For $r>\delta x$, the plaquettes no longer share common links, thus all correlations statistically vanish.
The full correlation function on the lattice is then given as a combination of \eqref{eq:corr_func_0} and \eqref{eq:corr_func_dx}:
\begin{equation}
    \label{eq:correlation_function_analytic_simple}
    \xi_{\rm iso}(r) \simeq \frac{3}{\delta x^4}\Big[\delta_{r,0}-\frac{1}{6}\delta_{r,\delta x}\Big]\,.
\end{equation}
We may observe that the relative numerical factor between the terms is purely due to the averaging over independent directions (1 in case of $r=0$ and 6 in case of $r=\delta x)$.
Using the definition of the set $\mathscr{P}(r)$ from \eqref{eq:P_def}, the isotropic correlation function may be written in a concise way as
\begin{equation}
    \xi_{\rm iso}(r)\simeq\frac{3}{\delta x^4|\mathscr{P}(r)|}\big(\delta_{r,0}-\delta_{r,\delta x}\big)\,.
\end{equation}

\begin{figure}[t]
    \centering
    \includegraphics[width=\linewidth]{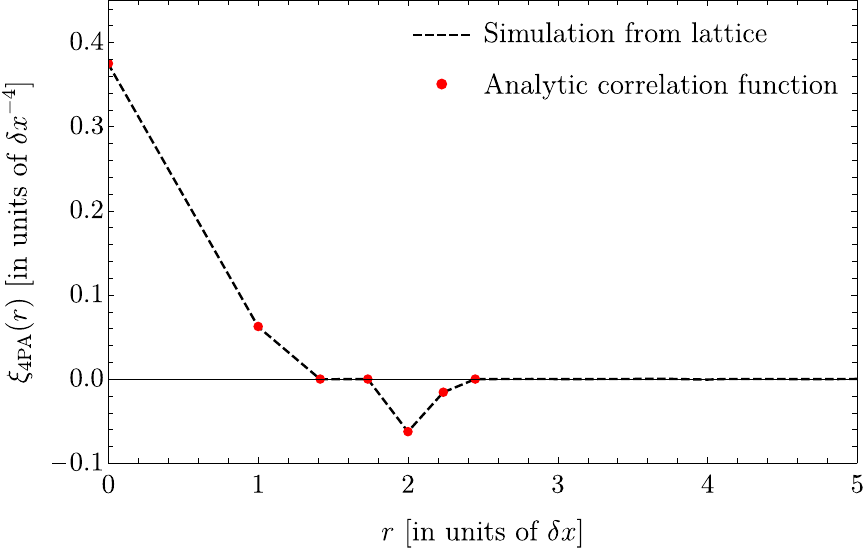}
    \caption{Comparison between the simulated correlation function on an $N=64$ lattice ({\it black, dashed}) and the analytic values ({\it red dots}) in the 4-plaquette averaged magnetic field case. 
    Simulation data points joined only for presentation.}
    \label{fig:correlation_compare_4PA}
\end{figure}

One can calculate averaged magnetic fields by considering larger than single-plaquette line integrals in \eqref{eq:B_definition}.
This, however, leads to correlations over larger distances.
In particular, in ref.~\cite{Vachaspati:2024vbw} a 4PA lattice was used in the main results.
The above presented algorithm to determine the correlation function can be identically applied to this case as well, with correlations persisting up to $r=\sqrt5\delta x$, \ie, where the enlarged plaquettes share only one common and anti-parallel link.
One can show that in this case the correlation function is
\begin{equation}
\label{eq:correlation_function_analytic_4PA}
\begin{aligned}
    \xi_{\rm iso}^{\rm (4PA)}(r)\simeq\frac{3}{8\delta x^4}&\Big[\delta_{r,0} + \frac{1}{6}\delta_{r,\delta x}-\frac{1}{6}\delta_{r,2\delta x}-\frac{1}{24}\delta_{r,\sqrt{5}\delta x}\Big]\\
    \equiv\frac{3}{8\delta x^4|\mathscr{P}(r)|}&\big(\delta_{r,0} + \delta_{r,\delta x}-\delta_{r,2\delta x}-\delta_{r,\sqrt{5}\delta x}\big)\,.
\end{aligned}
\end{equation}
If we redefine the lattice parameters such that the physical size of the 4 plaquettes is the same as the single plaquette in the original calculation, $\delta x\to \delta x/2$, then the $r=0$ and $r=\delta x$ terms match up with those in \eqref{eq:correlation_function_analytic_simple} up to a factor of 2 which is a consequence of having twice as many links in the 4PA case.
Fig.~\ref{fig:correlation_compare_4PA} shows a comparison of the analytic result \eqref{eq:correlation_function_analytic_4PA} with a lattice simulation with $N=64$.
The two match perfectly, even though the lattice is relatively small.
This confirms the result of the analytic calculation and confirms the validity of the large $N$ limit already at $N=64$.

\section{Expansion of the energy spectrum}
\label{app:expansion_of_the_energy_spectrum}

The power spectrum in \eqref{eq:b2_lattice} depends on the lattice wave-number vector $\bk_{m_x,m_y,m_z}=k_1(m_x,m_y,m_z)^{\rm T},$ where $0\leq m_x< N/2$ are integer numbers.
A similar, but lengthier sum of cosine functions appears for the 4PA magnetic field case.
We would like to use isotropy and average over the contributions from discrete wave-number vectors with the same norm.
This is most easily done by assuming a continuous spectrum, which is a reasonable approximation when $N\gg 1$.
To do so, we expand the cosine functions appearing in the expressions for the power spectrum.
The result is a sum involving monomials of the form $m_x^{2a}m_y^{2b}m_z^{2c}$ with $a,b,c$ non-negative being integers.
In the continuum we change $\{m_x,m_y,m_z\}\to\{x,y,z\}$ and we use 
\begin{equation}
    \label{eq:avg_formula}
    \langle x^{2a}y^{2b}z^{2c}\rangle=r^{2(a+b+c)}\frac{(2a-1)!!(2b-1)!!(2c-1)!!}{(2a+2b+2c+1)!!}\,,
\end{equation}
where $!!$ denotes the double factorial and $r^2=x^2+y^2+z^2$.

For the lattice result in \eqref{eq:b2_lattice} the general averaging formula of \eqref{eq:avg_formula} is not necessary as no mixed terms appear.
In fact, after Taylor-expanding the cosine functions and applying the averaging formula, the expression can be fully summed and we find for $k\leq k_{\rm Ny}$:
\begin{equation}
    \label{eq:b2_analytic_lattice_simple}
    \frac{\langle|\bb(k)|^2\rangle}{\delta x^2N^3}=3\big(1-j_0(k\delta x)\big)\,,
\end{equation}
where $j_0(kr)\equiv\sin(k r)/(kr)$ is the spherical Bessel function of order 0.
For $k>k_{\rm Ny}$ the results from \eqref{eq:b2_lattice} and \eqref{eq:b2_analytic_lattice_simple} differ as the analytic average implicitly assumes a spherical volume, while the lattice is cubic: in Fourier space the radius of the largest sphere that fits in the lattice is $k_{\rm Ny}$, consequently this is the largest wave-number for which the formula is applicable.

The energy spectrum is obtained via substitution of \eqref{eq:b2_analytic_lattice_simple} into \eqref{eq:energy_spectrum_def} (for $k<k_{\rm Ny}$):
\begin{equation}
    8\pi^2\frac{E_{\rm M}(k)}{k^4} = \frac{6}{k^2\delta x}\big(1-j_0(k\delta x)\big)\,.
\end{equation}
This energy spectrum is not compatible with physical expectations as $\rd E_{\rm M}(k)/\rd k=0$ has only one solution at $k_*=0$, \ie, the spectrum derived has no spectral peak.

In case of the 4PA magnetic fields, we can obtain the energy spectrum in an identical way,
\begin{equation}
\label{eq:b2_analytic_lattice_4PA}
\begin{aligned}
    &8\pi^2 \frac{E_{\rm M,4PA}(k)}{k^4} \\
    &= \frac{3}{4k^2\delta x} \big(1+j_0(k\delta x)-j_0(2k\delta x)-j_0(\sqrt{5}k\delta x)\big)\,.
\end{aligned}
\end{equation}
For small wave-numbers, the energy spectra in \eqref{eq:b2_analytic_lattice_simple} and \eqref{eq:b2_analytic_lattice_4PA} are identical if one uses $k_{\rm 4PA}=k/2$ to account for the artificially stretched space in the 4PA case.
In contrast with the original lattice spectrum, the 4PA spectrum resolves the spectral peak which lies at the numerical value $k_{\rm 4PA}\delta x\equiv 2k\delta x\approx 2.655$, which corresponds to a peak wavelength slightly above the correlation length, $\lambda_*\approx1.183\,\delta x$.

We mention here that the averaged energy spectrum, such as that of \eqref{eq:b2_analytic_lattice_4PA}, is directly available from the correlation function through the inverse of the relation given in \eqref{eq:EM_xi_relation}.
For a continuously defined correlation function $\tilde\xi(r)$ the relation is:
\begin{equation}
    \label{eq:EM_xi_relation_inverted}
    E_{\rm M}(k)=\frac{k^2}{\pi}\int_0^\infty\rd r\, r^2\tilde\xi(r)j_0(kr)\,.
\end{equation}
The lattice gives $\xi_{\rm iso}(r)$ that is only defined for discrete $r$ values allowed on the lattice. 
As mentioned below \eqref{eq:corr_func_simple_lat}, the isotropic correlation function $\xi_{\rm iso}(r)$ may be analytically continued for any $r$ by using
\begin{equation}
    |\mathscr{P}(r)|\to 4\pi\Big(\frac{r}{\delta x}\Big)^2\,,\quad\delta_{r,a}\to\delta x\,\delta(r-a)\,.
\end{equation}
The former is the generalization of the number of lattice points in a spherical shell of radius $r$, the latter follows from the definition of the Dirac-delta distribution.
For example, in the specific case of the 4PA correlation function, given in \eqref{eq:correlation_function_analytic_4PA}, the continuous version is:
\begin{equation}
\label{eq:xi_4PA_tilde}
\begin{aligned}
    &\tilde\xi_{\rm 4PA}(r) \\
    &= \frac{3\big[\delta(r)+\delta(r-\delta x)-\delta (r-2\delta x)-\delta(r-\sqrt{5}\delta x)\big]}{32\pi\delta x\, r^2}\,.
\end{aligned}
\end{equation}
Substitution of $\tilde\xi_{\rm 4PA}(r)$ into \eqref{eq:EM_xi_relation_inverted} yields the same energy spectrum obtained from the lattice calculation in \eqref{eq:b2_analytic_lattice_4PA}.

\section{Generalized coarse-grain algorithms}
\label{app:gen_algo}

The 4PA may be generalized to $n^2$PA algorithms.
The correlation function remains a sum of Dirac-delta distributions.
In general the $n^2$PA-type lattice correlation functions are of the form
\begin{equation}
\label{eq:Correlation_function_n2PA}
\begin{aligned}
    4\pi r^2\tilde\xi_{n^2{\rm PA}}(r)=\frac{3}{n^4\delta x}\bigg[ n\delta(r)+2\sum_{i=1}^{n-1}(n-i)\delta(r-i\delta x) \\
    -n\delta(r-n\delta x)-2\sum_{i=1}^{n-1}(n-i)\delta\big(r-\sqrt{n^2+i^2}\delta x\big)\bigg]\,.
\end{aligned}
\end{equation}
The sums are understood to contribute only for $n\ge 2$.
The number of positive and negative coefficients remain equal for any $n$ and these coefficients sum to zero, \ie, the following identity holds for any $n\geq 1$:
\begin{equation}
    \label{eq:conditions}
    \int_0^\infty \rd r \,r^2\tilde\xi_{n^2{\rm PA}}(r)=0\,.
\end{equation}
This is expected behavior for an isotropic correlation function as discussed later in app.~\ref{app:causality}.
If we expand the spherical Bessel function for small wave-numbers in the definition of the energy spectrum given in  \eqref{eq:EM_xi_relation_inverted}, we find that the identity given in \eqref{eq:conditions} ensures that the $k^2$-term in the expansion vanishes.
For completeness, we finally give the energy spectrum corresponding to \eqref{eq:Correlation_function_n2PA}:
\begin{equation}
\begin{aligned}
    &8\pi^2\frac{E_{{\rm M},n^2{\rm PA}}(k)}{k^4}=\frac{6}{n^3\delta x\,k^2}\Big[1+2\sum_{i=1}^{n-1}\frac{n-i}{n}j_0(i\delta x\,k)\\
    &\quad-j_0(n\delta x\,k)-2\sum_{i=1}^{n-1}\frac{n-i}{n}j_0(\delta x\sqrt{n^2+i^2}\, k)\Big]\,.
\end{aligned}
\end{equation}
It is simple to check that $E_{{\rm M},n^2{\rm PA}}(k)\propto k^4$ for any $n\geq 1$.

\section{Admissible spatial correlation functions}
\label{app:spat_corr}

We assume that the correlation function is exponentially small for large distances (causality condition) and has an arbitrary polynomial structure for low distances.
In general, the expected form is
\begin{equation}
    \label{eq:Cr_polynom_times_exp}
    \mathcal{C}(r)=\sum_{n=0}^{n_{\rm max}}p_nr^n\,\exp(-ar^2)\,,
\end{equation}
where $\{p_n\}$, $n_{\rm max}$, and $a$ are parameters of the spectra.
For convergence one requires $a>0$, while the polynomial is an even function of $r$, \ie, $p_{2n+1}=0$.
The energy spectrum is obtained via a Hankel transform, where the sum and the integral can be exchanged as the integrals are all finite due to the exponential suppression in $\mathcal{C}(r)$.
With $q=kr$ we find
\begin{equation}
    \mathcal{E}(k)=\frac{1}{\pi}\sum_{n=0}^{n_{\rm max}} p_nk^{-(n+1)}\int_0^\infty\!\!\rd q\,q^{n+2}\exp(-a q^2/k^2)j_0(q)\,.
\end{equation}
The integrals are analytic and convergent for any $n$, they can be expressed in terms of the confluent hypergeometric function of the first kind $_1{\rm F}_1(a;b;z)$:
\begin{equation}
    \mathcal{E}(k)=\frac{1}{2\pi}\sum_{n=0}^{n_{\rm max}} p_n\frac{k^2}{a^{m_n}}\Gamma(m_n){_1{\rm F}_1}\Big(m_n;m_0;-\frac{k^2}{4a}\Big)\,,
\end{equation}
where $m_n=(3+n)/2$.
The hypergeometric functions for positive and even $n$ have the form of a polynomial times an exponential function, mirroring the structure of the spatial correlation function.
The explicit expression, though not straightforward to derive \footnote{The main point of the derivation is to use Kummer's transformation for the hypergeometric function which factors out an exponential from the sum and terminates the originally infinite series.
The remaining steps are essentially trivial.}, is:
\begin{equation}
\begin{aligned}
    {_1{\rm F}_1}&\Big(m_n;m_0;-\frac{k^2}{4a}\Big) = \\
    &\sum_{j=0}^{n/2}(-1)^j\frac{k^{2j}}{(2a)^j}\binom{n/2}{j}\frac{1}{(2j+1)!!}\exp\Big(-\frac{k^2}{4a}\Big)\,.
\end{aligned}
\end{equation}
In order to have an energy spectrum that is at least $\propto k^4$, one requires that the $j=0$ terms of the hypergeometric function vanish.
This requirement gives a simple condition for the admissible polynomials in the spatial correlation function,
\begin{equation}
\label{eq:1F1_constraint}
    \sum_{n=0}^{n_{\rm max}}p_n a^{-n/2}\,\Gamma\Big(\frac{3+n}{2}\Big)=0\,.
\end{equation}

As an example, let us check the phenomenological energy spectrum from Sec.~\ref{sec:magnetic_field_continuous}.
There we introduced a $k^4$ energy spectrum in \eqref{eq:power_spectrum} and the corresponding correlation function in \eqref{eq:Cn_def}.
With the energy spectrum $\mathcal{E}^{(2)}(k)= (b_4k^4+b_6k^6)\exp(-2k^2/\sigma^2)$ the spatial correlation function becomes
\begin{equation}
    \mathcal{C}^{(2)}(r)\propto
    \big(p_0+p_2r^2+p_4r^4\big)\exp\Big(-\frac{1}{8}\sigma^2r^2\Big)\,,
\end{equation}
where $p_0=192b_4+240\sigma^2b_6$, $p_2=-(16\sigma^2b_4+40\sigma^4 b_6)$, and $p_4=b_6\sigma^6$.
This form mirrors that of \eqref{eq:Cr_polynom_times_exp} with $n_{\rm max}=4$ and $a=\sigma^2/8$.
For these parameters the constraint in \eqref{eq:1F1_constraint} becomes
\begin{equation}
    p_0+\frac{12}{\sigma^2}p_2+\frac{240}{\sigma^4}p_4=0\,,
\end{equation}
which is satisfied with the above values, confirming the correctness of our formulae.

\section{Causality and the energy spectrum}
\label{app:causality}

In this section we intend to clarify the relationship between the causality argument of \cite{Durrer:2003ja} and the integral condition on the spatial correlation function in \eqref{eq:k4_conditions}.
We consider a vector field $\bv(\bx)$ that is statistically isotropic and bounded, \ie, $|\bv(\bx)|\leq a<\infty$.
Isotropy implies that for each component of $\bv(\bx)$ the expectation value vanishes, $\langle v_i(\bx)\rangle=0$.
Using the ergodic principle we write this as a volume integral,
\begin{equation}
\label{eq:vi_expectation_value_0}
    \langle v_i(\bx)\rangle = \lim_{V\to \infty}\frac{1}{V}\int_V\rd ^3x \,v_i(\bx)=0\,.
\end{equation}
From boundedness of $\bv(\bx)$ it follows that each component is also absolutely integrable:
\begin{equation}
\label{eq:Fubini_condition}
    \lim_{V\to\infty}\frac{1}{V}\int_V\rd^3x\,|v_i(\bx)|\leq a<\infty\,.
\end{equation}

Due to isotropy, the spatial correlation function is expressed as $\mathcal{C}(\br)=3\mathcal{C}_i(\br)$, where $\mathcal{C}_i(\br) = \langle v_i(\bx)v_i(\bx+\br)\rangle$ is the correlation function for a single component of the vector field.
To proceed we calculate the expectation value of the correlation function:
\begin{equation}
\label{eq:Ci_expectation_value_0}
\begin{aligned}
    \langle \mathcal{C}_i(\br)\rangle&=\lim_{V\to\infty}\frac{1}{V^2}\int_V\rd^3 r\int_V\rd^3x\,v_i(\bx)v_i(\bx+\br) \\
    &=\lim_{V\to\infty}\frac{1}{V^2}\int_V\rd^3x \,v_i(\bx)\int_V\rd^3r\, v_i(\bx+\br)=0\,.
\end{aligned}
\end{equation}
In the second line we exchanged the order of integration, which is allowed by Fubini's theorem based on \eqref{eq:Fubini_condition}.
Then the $r$-integral is evaluated using \eqref{eq:vi_expectation_value_0} with an appropriate variable-shift.
Exploiting isotropy, the vanishing expectation value in \eqref{eq:Ci_expectation_value_0} is equivalent to the condition we introduced in \eqref{eq:k4_conditions},
\begin{equation}
\label{eq:integral_condition_derived}
    \int_0^\infty\rd r\,r^2\mathcal{C}_i(r)=0\,.
\end{equation}
This statement follows from isotropy and the boundedness of $v_i(\bx)$, or equivalently, from $v_i(\bx)$ being absolutely integrable, as shown in \eqref{eq:Fubini_condition}.

In order to connect \eqref{eq:integral_condition_derived} to the power spectrum $P(k)$, we need to further assume that the correlation function is exponentially small for large separations (\ie, we must introduce the causality condition). 
Then in the definition of the power spectrum we can Taylor expand the spherical Bessel function and exchange the integration and summation:
\begin{equation}
\begin{aligned}
    P(k) &\propto \int_0^\infty\rd r\,r^2C(r)j_0(kr) \\
    &= \sum_{n=0}^\infty\frac{(-1)^n}{(2n+1)!} k^{2n}\int_0^\infty\rd r\,r^{2+2n}C(r)\,.
\end{aligned}
\end{equation}
For any $n$ the integral is finite due to the exponential suppression assumed to be present in $C(r)$.
Clearly, the leading order ($n=0$) integral vanishes due to \eqref{eq:integral_condition_derived}, thus the power spectrum scales at least as $P(k)\sim k^2$.
We also mention that the expansion of $j_0(x)$ around $x=0$ has a convergence radius of $R_{\rm c}\to \infty$, so the above result is valid for all $k$.

The condition we derived in \eqref{eq:integral_condition_derived} directly followed from an isotropic and bounded vector field.
The same relation is obtained by considering a causal correlation function.
However, from \eqref{eq:integral_condition_derived} the scaling of the power spectrum does not follow unless causality is enforced.
This shows that \eqref{eq:integral_condition_derived} is simply a consistency equation with the causality argument and not a separate proof for the scaling of the power spectrum.

\section{Evolution of the spectra}
\label{app:evolution}

An important general detail of the energy spectrum is that isotropy of the underlying magnetic field and the causality of the spatial correlation function enforce that it should have a convergent expansion involving even powers of $k$,
\begin{equation}
    \label{eq:Ek_expansion_general}
    \mathcal{E}(k) = k^4\sum_{n=0}^\infty p_nk^{2n}\,,\quad p_n\in\mathbb{R}\,,
\end{equation}
such that the energy spectrum should be finite and integrable, resulting in $\mathcal{E}(k\to\infty)=0$.
This limit is then only possible if the expansion has infinitely many terms.
For example, the empirical formula used in \eqref{eq:power_spectrum} satisfies these constraints as the appearance of the exponential suppression requires the presence of an infinite expansion in terms of $k^{2n}$.

The integro-differential equations governing the evolution of a generic energy spectrum were derived from the MHD equations under assumptions of homogeneity, isotropy, and large conductivity in Refs.~\cite{Saveliev:2012ea,Saveliev:2013uva}.
As the evolution equations respect isotropy and causality, it follows that at every instant the spectrum should also be of the same causal and isotropic form in \eqref{eq:Ek_expansion_general}, in particular it should behave as $\propto k^4$ for small wavenumbers.
% The original integro-differential equations exhibited singularities in the integral kernels when the integration momentum coincided with the independent momentum variable.
In Refs.~\cite{Saveliev:2012ea,Saveliev:2013uva} the authors derived a set of coupled integro-differential equations that exhibited singularities in the integral kernels when the integration momentum coincided with the independent momentum variable ($\bk$ and $\bq$ below, respectively).
These singularities can be entirely removed by using the $\propto k^4$ infrared scaling of the energy spectrum. 
Consequently, compared to Refs.~\cite{Saveliev:2012ea,Saveliev:2013uva}, we modify their definitions of the magnetic ($M_k$) and kinetic spectral densities ($U_k$) by factoring out the causal scaling $k^4$ as
\begin{subequations}
\label{eq:Mq_Uq_redefs}
\begin{align}
    &M_k=k^4\langle|\bb(\bk)|^2\rangle/2\rho\equiv k^4\mathscr{M}_q\,, \\
    &U_k=2\pi k^4\langle|{\bf v}(\bk)|^2\rangle\equiv k^4\mathscr{U}_q\,,
\end{align}
\end{subequations}
so that $\mathscr{M}_k,\mathscr{U}_k\sim {\rm const.}$ for small wavenumbers, c.f., eq.~\eqref{eq:Ek_expansion_general}.
Here ${\bf v}$ is the velocity field of the fluid and $\rho$ is its mass density.
With these definitions the averaged evolution of the spectra are as follows.
In the non-helical case the evolution equation for the magnetic spectral density ($\mathscr{M}_q$) is
\begin{subequations}
\label{eq:dMq_dlna}
\begin{align}
    \!\!\!\frac{H_0}{a}\frac{\partial\mathscr{M}_q}{\partial\ln a}=\tau_c\int_{\mathbb{R}}\rd \ln k\,k^5\bigg[-\frac{2q^2}{3}\mathscr{M}_q \big( \mathscr{U}_k +2\mathscr{M}_k\big) \\
    +\frac{1}{2}\int_{-1}^1\rd x \,(1-x^2) (q^2+k^2-qkx) \mathscr{M}_k\mathscr{U}_{k_1}\bigg] \\
    -2\eta q^2\mathscr{M}_q\,,
\end{align}
\end{subequations}
while for the kinetic spectral density ($\mathscr{U}_q$) it reads:
\begin{subequations}
\label{eq:dUq_dlna}
\begin{align}
    \frac{H_0}{a}\frac{\partial\mathscr{U}_q}{\partial\ln a}&=\tau_c\int_{\mathbb{R}}\rd \ln k\,k^5\bigg\{-\frac{2q^2}{3}\mathscr{U}_q \big( \mathscr{U}_k +\mathscr{M}_k\big) \\
    +\int_{-1}^1&\rd x \,\frac{k}{4} \bigg[\bigg(k(1-x^2)+\frac{2k_1^2x}{q}\bigg)\mathscr{M}_k\mathscr{M}_{k_1}\,\\
    &\qquad\quad~+(3k-qx)(1-x^2)\mathscr{U}_k\mathscr{U}_{k_1}\bigg]\bigg\} \\
    &-2\nu q^2\mathscr{U}_q\,.
\end{align}
\end{subequations}
In either case, $\bk_1 = \bk-{\bf q}$ with the angle between $\bk$ and ${\bf q}$ given by $x\equiv\cos\theta$, and $\tau_c\simeq a/H_0$ is the de-correlation time.
In the original set of equations, inverse powers of $k_1$ appeared in the integrand, leading to singular behavior at $|\bk|\to|\bq|$ with $\cos\theta\to 1$.
The modified, explicitly causal set of equations in \eqref{eq:dMq_dlna}-\eqref{eq:dUq_dlna} are well-behaved and do not exhibit any singular behavior in the integration variables.

%\bibliography{apssamp}
%

\end{document}